# Non-Einsteinian Viscosity Behavior in Plasma-Functionalized Graphene Nanoflake Nanofluids and their Effect on the Dynamic Viscosity of Methane Hydrate Systems

*Adam McElligott, André Guerra, Chong Yang Du, Alejandro D. Rey, Jean-Luc Meunier, Phillip Servio\**

Department of Chemical Engineering, McGill University, Montreal, Quebec H3A 0C5, Canada

*phillip.servio@mcgill.ca

ABSTRACT

The dynamic viscosities of nanofluids containing oxygen-functionalized graphene nanoflakes (O-GNFs) were measured for concentrations ranging from 0.1 to 10 ppm under pressures from 0 to 30 MPag and temperatures from 0 to 10 °C. Water's viscosity dependence on temperature was not affected by the presence of O-GNFs though the effective viscosity of solution was reduced (termed non-Einsteinian viscosity) against common expectations. Hydrogen bond strength may have been reduced at the hydrophobic part of the O-GNF surface, whereas density fluctuations were enhanced. Therefore, larger sites of free volume may have formed, and weaker intermolecular interactions could allow for less-restricted diffusion into those sites, reducing the effective viscosity. The internal friction that would otherwise raise the solution viscosity could be overcome by these surface effects. Water's viscosity dependence on pressure was also not affected by O-GNFs, except at 10 ppm, where the shuttle effect may have increased the presence of hydrophobic methane bubbles in the solution. Under high pressure, the relative viscosity of the system remained non-Einsteinian at all temperatures except 2 °C. This may have been because the density anomaly of water was shifted to a colder temperature as the hydrogen bonding network was weaker. The phase transition from liquid to hydrate was identical to that of pure water, indicating that the presence of different stages of growth was not affected by the presence of O-GNF. However, the times to reach a maximum viscosity were faster in O-GNF systems compared to pure water. This said, the hydrate formation limitations inherent to the measurement system were not overcome by the presence of O-GNFs. The times to application-relevant viscosity values were maximized in the 1 ppm system at 49.75 % (200 mPa·s) and 31.93 % (500 mPa·s) faster than the baseline. Therefore, the presence of O-GNFs allowed for shorter times to desired viscosities and at lower driving forces than the baseline, improving the viability of the hydrate technologies to which they can be added.

1. INTRODUCTION

Methane is the primary component of natural gas, a major global source of non-renewable, primary energy.[1] It is also a potent greenhouse gas, and thus natural gas transport technologies that are under development now focus on further mitigating gas discharge to the atmosphere and increasing flow efficiency.[1] Specifically, transport is moving towards pressure vessels containing compressed or liquefied natural gas.[2] In addition to these, transport via methane gas hydrate has been increasingly examined due to the significant potential reductions in transportation costs and high storage capacities.[3]

Gas hydrates, also called clathrate hydrates, form when an inclusion molecule such as a gas becomes trapped in a thermodynamically stable cage made from the hydrogen bonds of water molecules. The crystal lattice of these non-stoichiometric compounds is stabilized through weak van der Waals forces from the enclosed molecule. Therefore, no chemical reaction or additional bonding occurs during the phase transition from liquid to solid.[4] Hydrate formation has three initial steps. The first is saturation, where the hydrate-forming molecule dissolves in the aqueous phase until a thermodynamic equilibrium concentration is reached.[4] This step is followed by induction, where hydrate nuclei form and dissolve in the now supersaturated solution until a critical radius is attained. At this point, the autocatalytic, exothermic growth step begins as hydrate growth becomes energetically favourable. Both heat and mass transfer processes can limit these initial steps, and their length can be affected significantly by altering the solution kinetics. Assuming there is sufficient gas present, growth continues at a linear rate until it becomes mass-limited (i.e., when there is insufficient water to continue forming cages), and the growth rate continuously slows until no more water is present. This study examines all these stages, though due to the significant formation driving forces, the initial two steps (saturation and induction) often occur in less than one second. Therefore, hydrate formation from the beginning of growth to near-solidification will be the focus of this study.[4]

Gas hydrates can form easy-to-transport compounds under less severe conditions than other strategies. For example, methane gas hydrates form at lower pressures than compressed natural gas and higher temperatures than liquefied natural gas.[5] Furthermore, hydrates have high gas storage potentials, efficient structural packing, and highly selective formation properties. These characteristics make them highly suitable for new technological applications.[4, 6-13] Therefore, other than energy transport, gas hydrates are being explored for novel separation processes, including flue gas cleaning, atmospheric $CO_2$ sequestration, and the concentration of fruit juices.[4, 6-12] To overcome heat or mass transfer limitations in these new processes, it is often necessary to optimize the gas hydrate formation properties via additives; commonly surfactants or nanoparticles. The nature of hydrate promotion can be either thermodynamic, shifting the three-phase equilibrium curve towards more favourable formation conditions, or kinetic, inducing the nucleation of stable hydrate clusters.[14] Examples of hydrate-promoting nanoparticle additives include metal nanoparticles, especially metal oxides, and carbon-based nanoparticles like graphenes and single- or multi-walled carbon nanotubes. Graphene nanoflakes (GNFs) are two-dimensional materials of stacked graphene planes that enhance the yields of methane and other gas hydrates.[15-20] Pure carbon GNFs are hydrophobic and require modification to remain stable in solution. Usually, chemical treatment or surfactant addition is required to avoid nanoparticle agglomeration and settling.[20, 21] Now, oxygen or nitrogen functional groups can be added onto GNF surfaces via thermal plasma decomposition and chemical functionalization processes imparting hydrophilicity onto the particles. Recent studies have shown that plasma-functionalized GNFs significantly improve hydrate formation, with oxygen functionalized GNFs (O-GNFs) nearly quadrupling methane hydrate growth rates.[21-23] However, while promotion is essential for improving hydrate technologies, it is insufficient to demonstrate their viability alone. Several new hydrate technologies propose using either semi-batch or, more often, continuous processes during full operation. Maintaining a flow state while hydrate formation is induced requires significant control of solution viscosity to reduce pumping requirements and avoid complete solidification. Therefore, characterizing these nanofluids' temporal evolution of viscosity under various thermodynamic (hydrate-forming) conditions is critical to governing the design and optimization of these processes.

This study characterizes the dynamic viscosity of methane hydrate formation in O-GNF nanofluids systems. The thermodynamic conditions will be combinations of temperatures from 0 to 10 °C and pressures from 0 to 30 MPag. Preceding reports have studied hydrate systems from pure water under the same temperature and pressure conditions. There have also been many recent studies on the viscosity of graphene and graphene oxide nanofluids.[24-35] This is the first time the viscosity of plasma-functionalized graphene has been measured in any system, including liquid and hydrate systems. Though other nanofluids have been analyzed at similar pressures, evaluating the viscosity of graphene nanofluids at high pressure is also entirely novel.[36-39] Additionally, this study is paired with another publication which used oxygen-functionalized multi-walled carbon nanotubes in similar systems.[40] A comparison between these two studies is found in that study and will not be further referenced here.

## 2. MATERIALS AND METHODS

### 2.1 Experimental Setup

The experiments analyzed in this study were performed in the system first developed by Guerra et al. (2022), depicted in simplified form in **Figure 1**. The primary measurement device was an Anton Paar MCR 302 rheometer. It was equipped with a high-pressure cell with a 40 MPag maximum pressure rating. The measurement geometry inserted into the cell was a double-gap geometry consisting of a double annulus space where the sample could be loaded. Rotational motion was induced in this geometry via a magnetic measurement head. Sample temperature was maintained by a Julabo F-32 chiller using a refrigerant fluid that was a 50/50 mixture of ethylene glycol and water by volume. The methane used in the experiments was purchased from MEGS and was of ultra-high (99.99%) purity, while the reverse osmosis (RO) water was produced with a 0.22 μm filter and had a maximum organic content of 10 ppb. Pressures up to 10 MPag were achievable using the gas cylinder alone, though higher pressures required gas compression prior to rheometer pressurization. A Schlumberger DBR high-pressure positive displacement pump, a mechanical piston system, was used to compress methane gas samples in the piston chamber to reach the pressures required for this study.

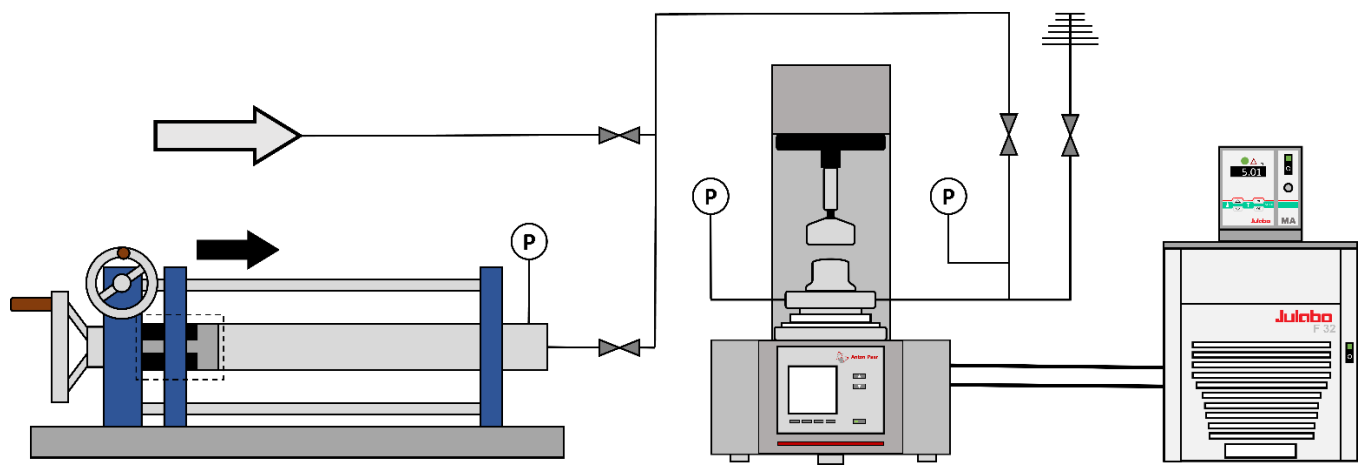

**Figure 1.** Simplified schematic of the experimental setup. It consisted of a mechanical piston system, an Anton Paar MCR 302 rheometer, and a Julabo F-32 chiller from left to right. The large grey arrow in the top left represents the air or methane gas inlet.

2.2 Characteristics of Oxygen-Functionalized GNFs

The O-GNFs described herein were produced and characterized in McGill University's Plasma Processing Laboratory prior to this study. Therefore, the fabrication process will be provided only briefly, and a focus will be placed on those aspects most relevant to the final characteristics of the nanoparticles, and so those having the greatest pertinence to the effects the O-GNFs engender in the system later. O-GNFs are produced in two stages. The first is a homogeneous nucleation process that creates critical carbon clusters from carbon vapours using a methane/nitrogen feed exposed to an argon plasma. This stage results in a pure powder of graphene sheets. In the second stage, the gas feed is changed to air. Active species from the oxygen in the air form and interact with the GNF surface, adding hydrophilic oxygenated functionalities such as carboxylic groups, hydroxides, or ether oxides to the surface. At the end of this stage, the O-GNFs have a thickness between 5 and 20 atomic layers, 10 on average, and planar dimensions of about 100 x 100 nm$^2$. The atomic composition of the surface is mostly carbon, though 14.2 % of the surface is oxygen.[41] This results in a stable nanofluid that does not require the presence of a surfactant to be perfectly dispersed and exhibits no agglomeration for periods of months to years. Further information regarding the production, functionalization, characterization, and imaging of the O-GNFs is found in Legrand et al. (2016).

2.3 Experimental Procedure

Prior to beginning a test run, 7.5 mL of the O-GNF solution was loaded into the well of the high-pressure cell. The measurement geometry was then inserted and used to close the cell. The sample's headspace was purged five times to eliminate the presence of air, one minute each time, using methane at a pressure of 1 MPag. Once the sample temperature was stable within 0.1 °C of the setpoint temperature, the measurement system of the rheometer was activated, and the cell was instantly charged to the test pressure with methane gas. This gas came either directly from a gas cylinder or the 500-cc piston chamber of the positive displacement pump. The rheometer ran at a constant 400 $s^{-1}$ shear rate regardless of condition, which is the recommended shear rate for double-gap measurement geometries in the case of low viscosity liquids like water. The shear rate needed to remain constant throughout a run so that temporal viscosity measurements during the different stages of hydrate formation were consistent and comparable. Therefore, the Newtonian or non-Newtonian nature of O-GNF nanofluids, which would require shear rate changes, is out of the scope of this study.

The pressures examined in this study were from 0 to 5 MPag going up by 1 MPa plus 10 to 30 MPag going up by 5 MPa; a total of 11 pressures. The temperatures examined were from 0 to 10 °C going up by 2 °C, totalling six temperatures and thus 66 conditions total when including the pressure range. These are the same conditions as Guerra et al. (2022), who examined methane hydrate formation in pure water systems.[24] That study will be used as a baseline for comparison with the results of the present study. The majority of these conditions are classified as hydrate-forming: they are found above the three-phase equilibrium line, so there is a positive driving force for hydrate formation.[42] **Figure 2** outlines all conditions examined in this study and denotes with an "x" which ones are expected to form hydrates. All hydrate-forming runs were given a maximum 90-minute period to begin hydrate formation, which was detected by a sharp increase in viscosity. It could equally be detected by a sudden rise in solution temperature. After formation began, viscosity measurements continued until the rheometer's set maximum torque limit (115 mN·m) was reached. This value occurs a few seconds before complete solidification of the sample, which was not allowed to arise as consistent solidification could significantly damage the ball bearings in the magnetic measurement head. Therefore, at this point, the rheometer would automatically stop collecting data, and the test run would end. For conditions not expected to form hydrates, measurement proceeded until a continuous ten-minute period of stable viscosity values was achieved. This ensured that any effects from temperature changes or gas dissolution at the start of the run were eliminated from the final reported viscosity value. The concentrations of the O-GNF solutions were 0.1, 1, and 10 ppm; all conditions were tested for each of these concentrations. Note that, as the viscosity of plasma-functionalized carbon nanofluids has never been measured, unpressurized runs of O-GNF nanofluids with concentrations of 0.1, 0.5, 1, 5, and 10 ppm by mass were also examined in the concentration range in an air atmosphere (i.e., with no methane present). This concentration range is rarely investigated in nanofluid viscosity studies and has never been investigated for carbon nanofluids, regardless of functionalization. The Anton Paar software RheoCompass v.1.25 was used for data collection, and data analysis was performed in MATLAB®.

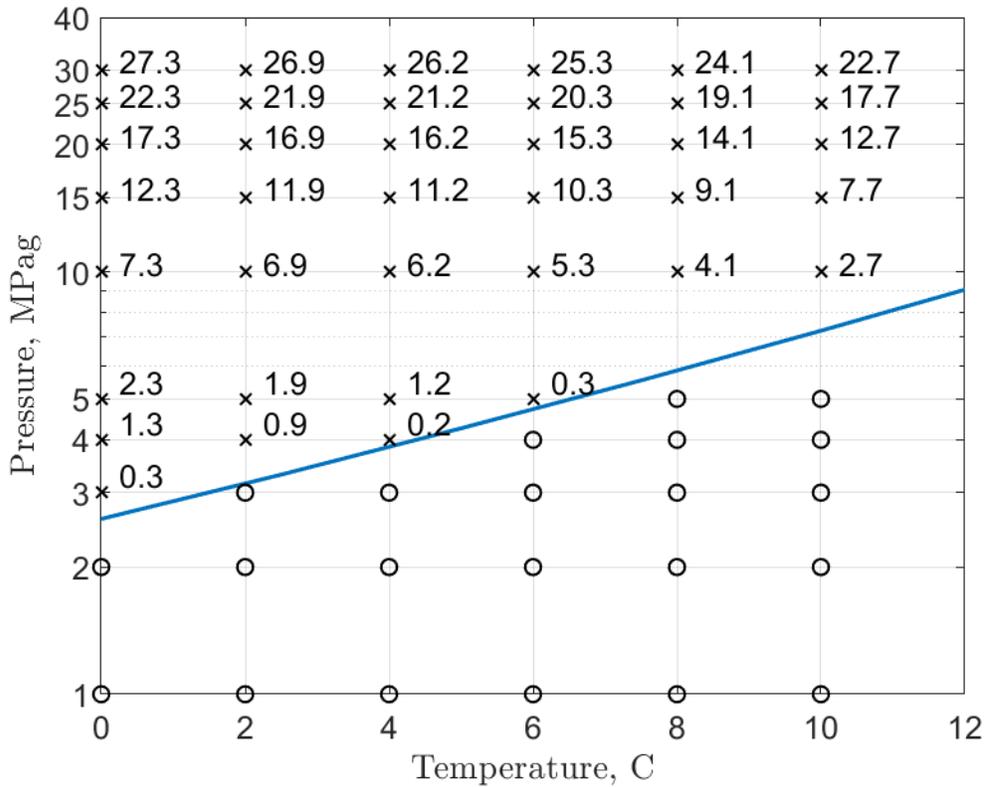

**Figure 2.** Pressure driving forces from the methane three-phase equilibrium curve[42], denoted by "x", and in the context of the temperature/pressure conditions examined for each concentration in this study. Non-hydrate-forming conditions are denoted by "o". Note that the six conditions at 0 MPag are not depicted as the vertical axis uses a log scale for clarity.

## 3. RESULTS AND DISCUSSION

### 3.1 Viscosity of O-GNF Nanofluids

The solution viscosity at atmospheric pressure (without the formation of hydrates or the presence of methane) was first examined for the O-GNF nanofluids to provide reference values to which the results of the later methane-pressurized tests would be compared. This is because the viscosity of plasma-functionalized graphene nanofluids, and the effects of their presence, have yet to be measured. Therefore, viscosity was measured for concentrations of 0.1, 0.5, 1, 5, and 10 ppm for each temperature condition (from 0 to 10 °C, increasing by 2 °C). The effects of concentration and temperature on viscosity were then investigated, as well as relative nanofluid viscosity compared to pure water.

### 3.1.1 Temperature Effects

The effects of temperature on O-GNF nanofluid viscosity are presented in **Figure 3**. The absolute viscosity of the solution decreases linearly with increasing temperature for all concentrations. Decreases in viscosity with temperature are commonly observed in most nanofluids, including graphene nanofluids, and over much greater concentration and temperature ranges than the ones presented in this study.[25-27, 30, 32-34, 43] However, it is often noted that these decreases largely depend on the properties of the base fluid and not on the presence of the nanoparticles themselves.[25, 44, 45] When a fluid is heated, higher energy is supplied to the fluid molecules.[46] This increases the fluid particles' random, Brownian motion (thermal movement) and average speed. In turn, both intermolecular adhesion forces and fluid particle-particle interactions are reduced.[45-47] Thus, the fluid's resistance to shearing is reduced, and viscosity decreases. It can also be noted that some nanoparticles are known to agglomerate with increasing temperature. However, this is uncommon for particles with high zeta potentials such as O-GNF,[41] and so is not predicted to affect the system's viscosity.[45] To provide further evidence that O-GNFs do not affect the temperature-viscosity relationship, the viscosity of the nanofluid relative to that of the base fluid is plotted in **Figure 3**b. It was found that there was essentially no change in relative viscosity as temperature increased at any concentration, which indicates that the nanoparticles do not affect how viscosity changes with temperature. Therefore, as the system temperature increases, the solution at any concentration in the range can be considered as the same dispersion of nanoparticles, simply in a lower-density liquid.

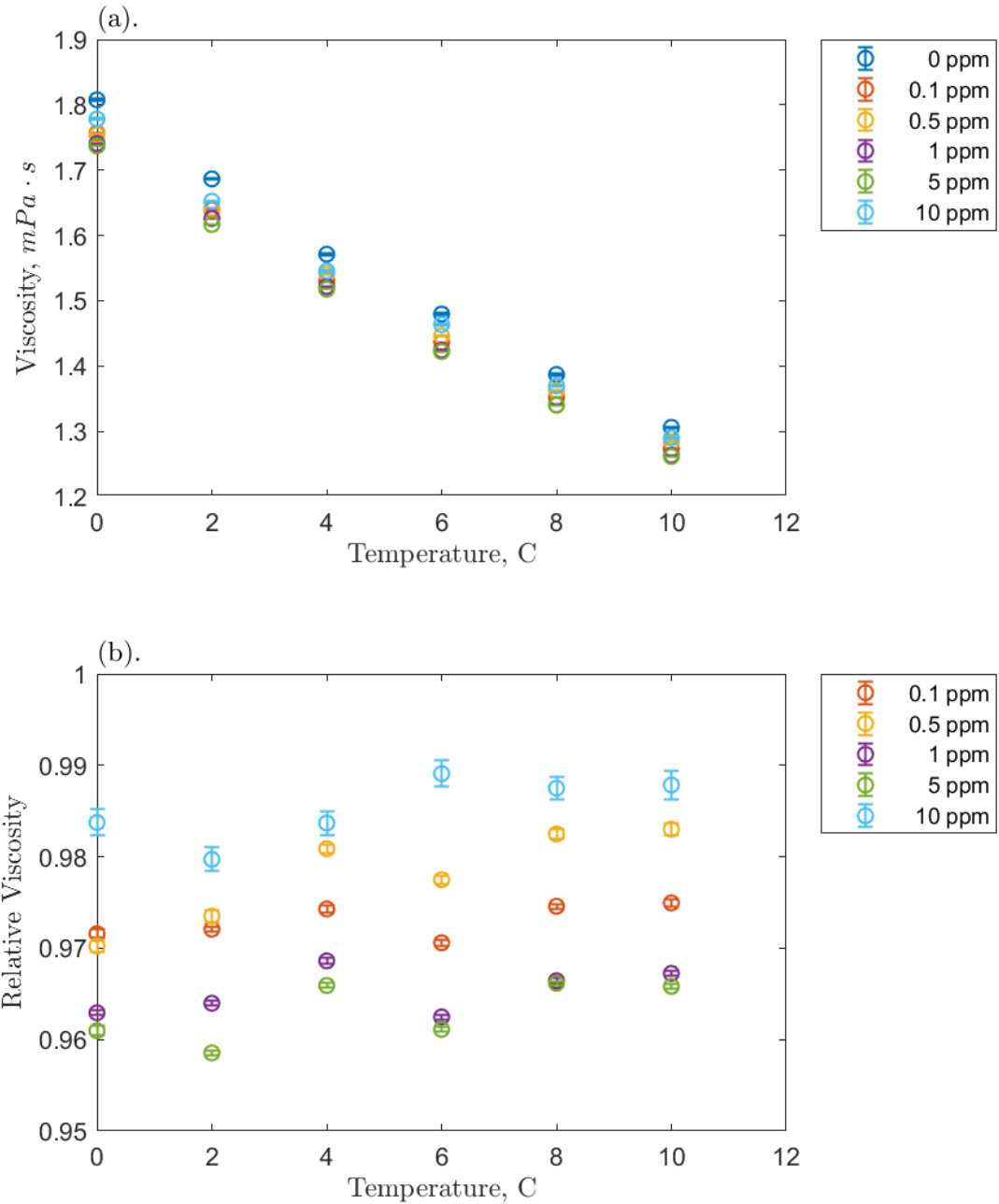

**Figure 3.** Effect of temperature on the O-GNF-water system viscosity. Part (a) is the absolute viscosity and part (b) is the relative viscosity compared to a pure water. The error bars are the 95% confidence intervals.

**Figure 3**b demonstrates a notable effect of O-GNFs in the solution: the relative viscosity, regardless of concentration or temperature, has a value of less than one. This means that the addition of the nanoparticle reduced the effective viscosity of the system, which is contrary to most studies examining nanofluid viscosity.[48] This includes graphene nanofluid studies, though it is notable that none have examined concentrations as low as those investigated here.[25-27, 29-35] The explanation for changes in the effective viscosity with the addition of solid particles is usually traced back to Einstein, who suggested that introducing very small rigid spheres to a fluid system increased its internal friction coefficient.[49] This study will term nanofluids that exhibit reduced viscosity as non-Einsteinian or NE nanofluids. Nanoparticles are primarily considered to enhance viscosity by increasing fluid resistance to movement, hindering the fluid particles' translational and rotational motion and thus creating higher frictional forces.[45, 50, 51] When more drag effects (strongly correlated to particle shape) are present, there is a greater dissipation of energy, which results in a higher effective viscosity.[46] The addition of solid particles also introduces particle-liquid interactions.[52] Notably, the attraction of counterions to the nanoparticle surface leads to the formation of an electrical double layer. When the double layers of different suspended particles interact, this creates a new electroviscous force, increasing the effective viscosity of the solution. However, non-Einsteinian behaviour has been observed previously, though usually in systems with metal oxide nanoparticles in glycol base fluids.[45, 53-56] This behaviour was ascribed to nanoparticle-liquid interactions that disturbed intermolecular hydrogen bonding. This kind of disturbance would be significantly effective in glycols, which have multiple -OH bonds.[55, 56] Additionally, some NE effects have been observed in water-based nanofluids, including some with carbon nanomaterials.[57, 58] It was suggested that this was caused by the lubricative effect of multi-walled carbon nanotubes (MWCNTs), though no further clarification was provided. This is not a sufficient

explanation as water, like glycols, also has a strong hydrogen bond-dependant viscosity and the previously cited studies on graphene nanofluids show increases in viscosity, including when water is used as a base fluid. Therefore, in the following section, a new hypothesis will be presented that offers the first brief though comprehensive conjecture on how non-Einsteinian effects could exist in systems such as the ones present in this study.

### 3.1.1.1 Hypothesis of Non-Einsteinian Viscosity in Water

This section aims to provide an initial hypothesis and research direction for the investigation of non-Einsteinian effects in nanofluids that goes well beyond previous attempts at explanation. It is important to note that the effects proposed here are on the molecular or nano scale and thus were not tested as part of this macro-scale study, though significant literature backing is provided to link these effects with those observed in this study. Computational modelling is required to test the hypothesis and is suggested as future work. The hypothesis can be stated as follows: at ultra-low concentrations, dispersed nanofluids with high specific surface areas may not provide sufficient internal friction in sheared water to increase the effective viscosity of the solution. Instead, hydrophobic interactions could disrupt local hydrogen-bonding networks, which would result in an effectively lower nanofluid viscosity. In addition, the presence of solid structures in the solution may enhance density fluctuations and improve their momentum flux in the shearing direction, further reducing viscosity.

As is well known, discussions of concentration-dependent viscosity effects usually begin with Einstein, who suggested that in a colloidal suspension of rigid spheres in a homogeneous liquid, solid particles increase the coefficient of internal friction in the fluid and thus the effective viscosity, as explained in the previous section. It was also assumed that the velocity components of the surface and adjacent liquid particle elements corresponded to each other (i.e. there was a no-slip condition).[49] It is proposed here that at ultra-low concentrations, viscosity-reducing surface effects are dominant over any drag effects induced by the presence of solid particles if the particle surface area is sufficiently high. As NE behaviour was observed at all concentrations investigated for O-GNF, an upper loading limit to these effects cannot be provided. It should also be noted that commonly used models for nanofluid viscosity, such as the Batchelor or Krieger-Dougherty equations, fail to predict non-Einsteinian viscosity values (i.e., at all the concentrations examined in this study, they predict no change in viscosity or, more precisely, a relative viscosity of 1.00).

When hydrophobic solids are dispersed in water, a repulsive solvation layer forms at the solid-liquid interface through hydrogen bonding which disrupts the hydrogen bond network.[56, 59] Hydrogen bonds are the dominant intermolecular force in water, an order of magnitude greater in strength than van der Waals forces.[60, 61] Reducing the attraction forces between water molecules at the solid surface would lower local density and viscosity. O-GNFs are considered hydrophilic in that they can be dispersed in water. However, the presence of ions that maintain the electric double layer likely does not enhance or break down hydrogen bonds.[62] Instead, the effect of the surface functional groups on viscosity may be that they result in a more stable suspension and maximize the hydrophobic surface area, approximately 85% of the total O-GNF surface area, for solid-liquid interaction. Additionally, the graphene used in this study has a very high specific surface area of 2630 $m^2$ $g^{-1}$. This could explain the observation by Cabaleiro et al. (2018), who found that graphene nanofluids exhibit lower effective viscosities with more hydrophilic graphene oxide sheets.[25] The no-slip assumption does not apply to hydrophobic graphene surfaces where there is reduced friction between water and the solid wall (slip) and effective repulsion of the water molecules, so lower viscosities again could be expected. Notably, water flow at the graphene surface has been measured to be significantly greater than predictions made using a no-slip condition.[63]

Viscosity can be free volume-limited, depending on the free space available for molecules to move. If more molecules are crowded in a small space, viscosity increases.[64] For diffusion to occur, a molecule must attain sufficient energy to overcome the local attraction forces of other molecules, and a local empty site of sufficient size must be available to which the molecule can diffuse.[65] The latter is governed by stresses generated in the fluid. Namely, local density fluctuations must be sufficient to open an ample enough space to permit displacement.[65] These density fluctuations give rise to intermolecular momentum transport and influence viscosity.[66] Higher local density fluctuations may occur at the nanoparticle surface, up to three atomic diameters away, inducing local free volume increases.[67, 68] Hydrophobic particles, through the formation of a solvation layer, may specifically lower the work required to create a cavity within the hydrogen bond network.[69] Furthermore, OH bonds that are not hydrogen-bonded to another oxygen are called free OH bonds. These bonds promote water mobility as the activation energy required to move to the closest equilibrium position is reduced. Jin et al. (2018) reported that the number of free OH bonds is increased at the graphene surface, which, in turn, enhances the flow of water molecules.[70] Therefore, at the solid-liquid interface, there may be a greater number of empty sites for water diffusion which are paired with weaker intermolecular attraction forces. When the available surface area is sufficiently high, these effects may overcome any internal friction effects at ultra-low concentrations and effectively lower viscosity.

Additionally, larger-scale shearing may play a role in reducing viscosity. The density fluctuations caused by shearing could be considered acoustic waves. The presence of nanoparticles with solid, crystalline structures may augment the propagation of acoustic waves as they travel more quickly through solids than liquids. Transverse acoustic waves may be amplified in this system and enhance momentum transport through collisions between water and the particle. Specifically, a wave with a certain pre-collision velocity has a much faster velocity after the collision as it travels through that nanoparticle. When this wave goes back to the liquid state, its velocity would be reduced, but it could now be greater than what it was prior to travelling through the solid. Therefore, a greater flux in the shear direction would result, the impedance to flow would be reduced, and the result would be an effective decrease in viscosity.[67] Lastly, it can be noted that shear-induced particle migration has been reported in nanofluid systems where particles move away from the walls of the geometry and the measured viscosity is lower.[71] However, it is unlikely that this would have a significant effect due to functional groups maintaining some distance between nanoparticles. Particle migration is also less likely for nanoparticles in the size range of O-GNFs.[71] In short, nanoparticles may enhance the free volume available to the liquid and lower the effective solution viscosity. However, these effects are only predominant at ultra-low concentrations where internal friction is minimized while the specific surface area is high.

### 3.1.2 Concentration Effects

With potential explanations of the viscosity effects in the presence of O-GNFs provided, the question remains as to how these effects change with different concentrations. Examining **Figure 3**b, the order from lowest to highest relative viscosity can be divided into the middle concentrations (1 and 5 ppm, around 0.96), the lowest concentrations (0.1 and 0.5 ppm, around 0.975), and finally, the highest concentration (10 ppm, about 0.985). All values were between 0.955 and 0.99. As mentioned, temperature did not significantly affect relative viscosity and can be removed from this part of the discussion. Some NE effects were likely present at the lowest concentrations as the relative viscosity was less than one. These effects became more significant as loading increased to 1 and then to 5 ppm as the relative viscosities were further reduced. It is noteworthy that increasing the loading continued to reduce viscosity as it could further indicate that internal friction, which would increase with loading, is not a significant factor at these concentrations. Lastly, there was an increase in relative viscosity at 10 ppm. This may be related to an increase in collisions from the higher density of nanoparticles, creating more microscale eddies and localized vortices within the fluid matrix. These would increase the effective system viscosity.[44] This kind of mean free path limitation has been suggested in previous graphene studies and is often predominant at higher concentrations.[21-23] It is likely that at even higher concentrations, as near as 20 ppm, the internal friction effects become dominant, and the relative viscosity becomes greater than one. In other words, the rise at 10 ppm may mark the beginning of the transition to the Einsteinian regime. However, there may still be many successive concentrations before this is reached that are beyond the range of this study.

### 3.2 Dynamic Viscosity of Methane Hydrate and O-GNF Systems

#### 3.2.1 Temperature and Pressure Effects

Building on the previous section, both pressure and temperature effects were measured in the same temperature range, now for O-GNF concentrations of 0.1, 1, and 10 ppm. Using isobars and isotherms, these effects are presented in  However, hydrate formation did not occur under these conditions, which will also be discussed in the next section.

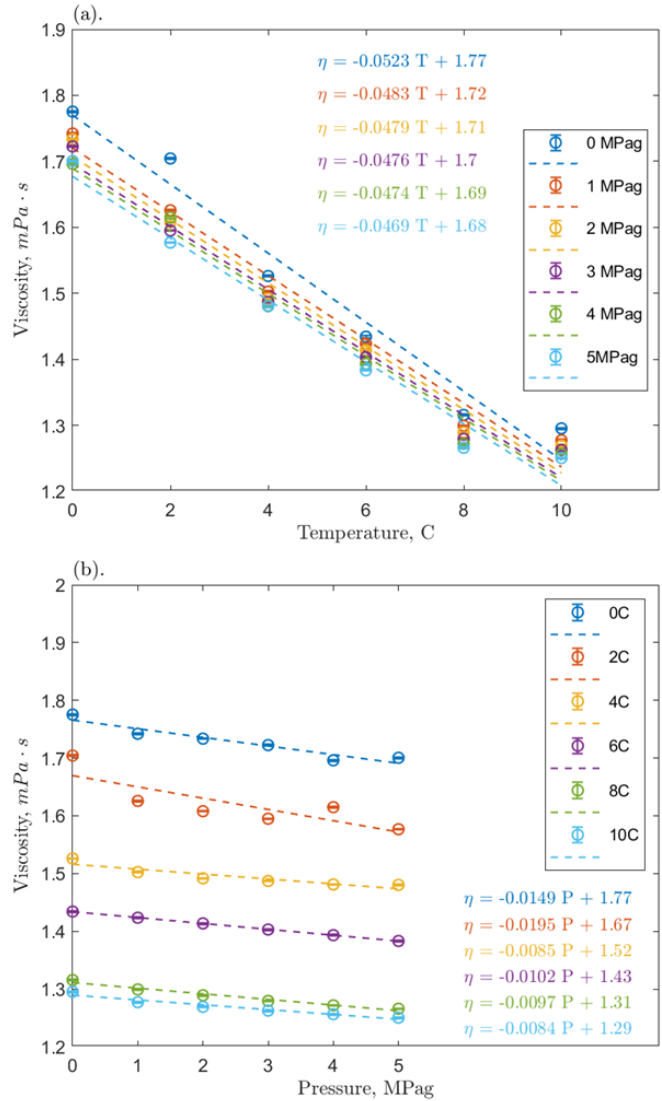

**Figure 4** for the 10 ppm O-GNF system. The figures for the other concentrations (0.1 and 1 ppm) can be found in the supplemental information. Each test condition consists of an average of 110 points taken over exactly 10 minutes at a constant shear rate of 400 s$^{-1}$. Constant liquid viscosity values in systems with pressures of 10 MPag and above could not be measured as hydrate formation was near-instantaneous; the viscosity profiles of those systems are found in the following section. Therefore, **Figure 4** only shows the 0 to 5 MPag measurements. By examining

**Figure 2**, eight hydrate-forming conditions should be present in this smaller range of pressures. However, hydrate formation did not occur under these conditions, which will also be discussed in the next section.

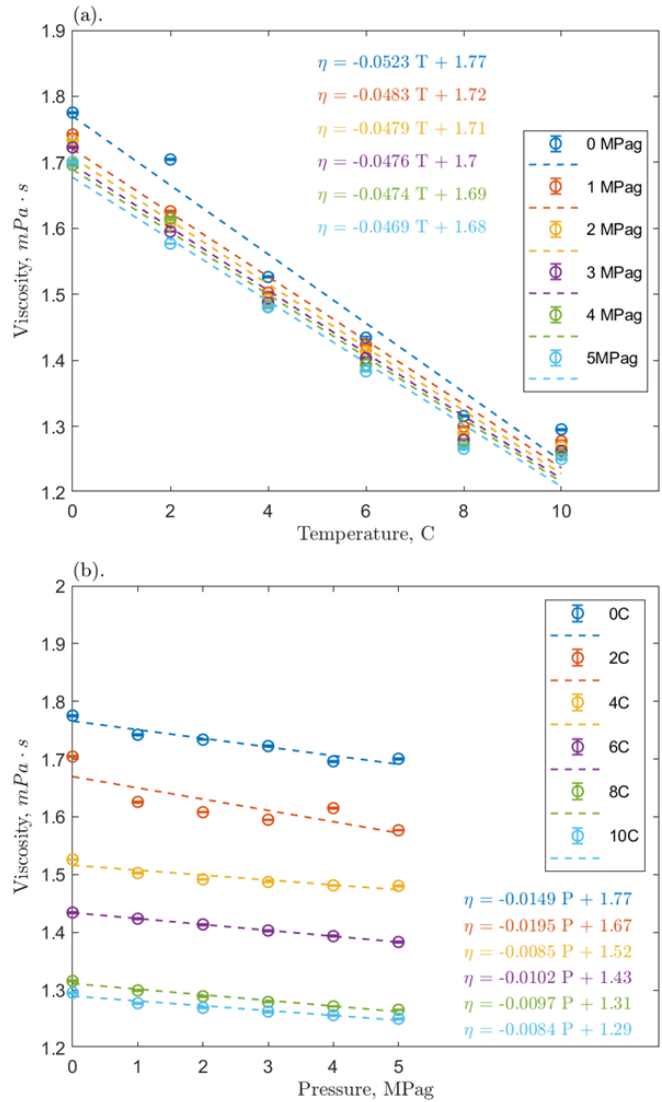

**Figure 4.** Viscosity effects in the 10 ppm O-GNF-methane-water system of (a) temperature and (b) pressure. The error bars are 95% confidence intervals and linear regressions are provided for each condition.

Using linear regressions, temperature and pressure effects were analyzed across the range. **Figure 4**a shows that as temperature increased, again nanofluid viscosity decreased. Compared to each other, the slopes are very similar regardless of pressure. The average slopes were -0.0502, -0.04, and -0.0484 mPa·s/°C for the concentrations of 0.1, 1, and 10 ppm O-GNF, respectively. Note that the pure water/methane baseline has an average slope of -0.0445 mPa·s/°C.[24] All slopes are again negative: the addition of O-GNF had no measured effect. Therefore, it is likely that O-GNFs do not have a significant effect on how the viscosity of water changes with temperature. This conclusion is consistent with other examinations of high-pressure nanofluid viscosity, where changes in viscosity changes with increasing temperature depended mostly on the properties of the base fluid.[37, 72] Some models, like the Vogel-Fulcher-Tammann-Hesse or Arrhenius equations have previously been used to characterize nanofluid viscosity's temperature dependence. However, the temperature range is relatively small; exponential equations are usually employed for ranges greater than 30 °C. Therefore, the fits here would still appear linear if either equation were applied and, to avoid overfitting, simpler linear equations are used. Briefly, however, the VTF equation is:

$$\frac{\eta}{\eta_0} = e^{\frac{A \cdot T_0}{T - T_0}}$$  Equation 1

Where $\frac{\eta}{\eta_0}$ is the relative viscosity, A and $T_0$ are fitting parameters, and T is the temperature in Kelvin.[43] The values of the fitting parameters were calculated but will not be provided here because overfitting may limit their physical significance. However, it can be mentioned that the values of the fitting parameters are constant for each concentration. This is consistent with the linear analysis and adds further evidence that the presence of O-GNF does not influence how viscosity changes with temperature.

**Figure 4**b shows that nanofluid viscosity did not change significantly with pressure. The slopes were alike across the temperature range, and the average slopes were -0.0082, -0.0034, and -0.0119 mPa·s/MPag for the concentrations of 0.1, 1, and 10 ppm O-GNF, respectively. Note that the pure water/methane baseline has an average slope of 0.00125 mPa·s/MPag.[24] All the slopes are all close to zero and this trend was not affected by the presence of O-GNFs. However, it is notable that the average slope at 10 ppm is negative and an order of magnitude greater than that for the baseline (noting that this was not the case for the 0.1 and 1 ppm conditions as seen in the Appendix and the slope is still about four times smaller than those for temperature dependence). The shuttle effect, where hydrophobic methane molecules are "shuttled" into the liquid bulk after adsorbing onto the hydrophobic portion of the O-GNF, may be responsible for this behavior.[73] Methane has previously been found to reduce the effective solution viscosity (through hydrophobic effects) when added to water.[24] If the O-GNFs supplied extra methane to the solution, there could be a decrease in the measured nanofluid viscosity. There have been previous studies where the addition of O-GNFs did not significantly change how much methane dissolved into the liquid phase of the system.[22] However, non-Einsteinian behavior may only depend on an accumulation of surface effects, so small amounts of methane could still impart a significant effect on the viscosity. Therefore, there may be a weak, negative effect of nanoparticle addition on the pressure-dependence of viscosity in this system. Previous high-pressure nanofluid viscosity studies have reported an increase in viscosity with pressure.[37, 72, 74] However, like temperature dependence, this may be more closely related to the properties of the base fluid, and the viscosity of water does not change significantly in this pressure range.

The relative viscosity of solution across the temperature and pressure range was then measured for each concentration and the results are found in **Figure 5**. As a previous part of this section demonstrated that viscosity depended only weakly on pressure, the values are averaged over 0 to 5 MPag for each concentration/temperature condition. Moreover, the base fluid is no longer pure water here, but instead water pressurized with methane (again averaged over the pressure range). This was done to eliminate the influence of methane, which can lowers the solution viscosity.

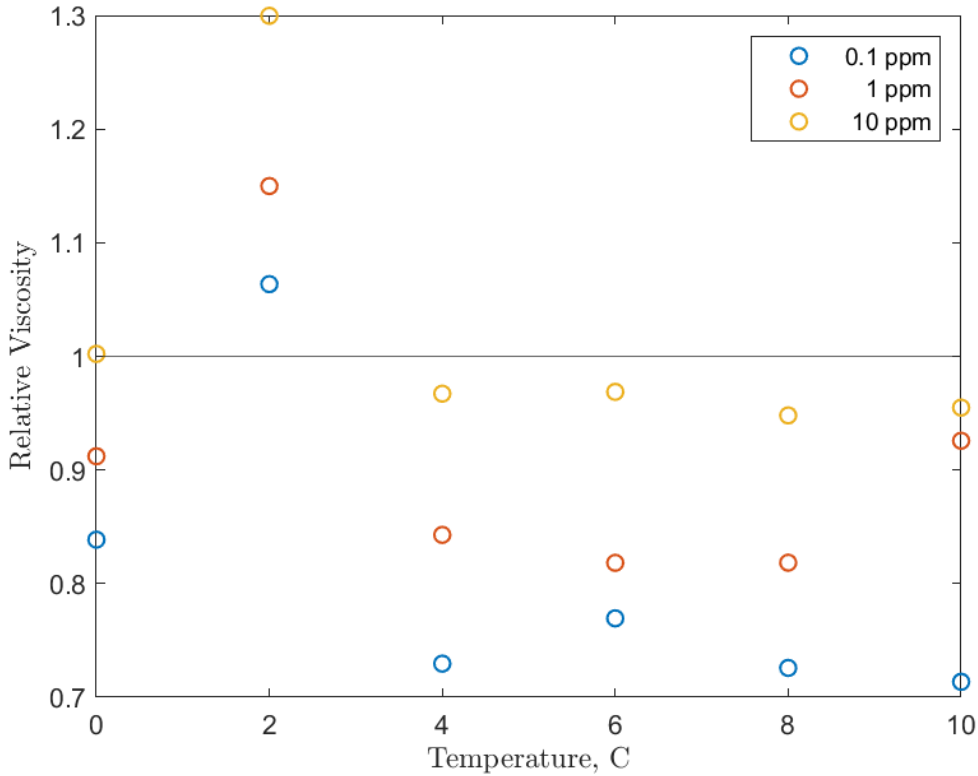

**Figure 5.** The effect of temperature on the relative viscosity of O-GNF-methane-water systems. Each value is an average over pressures of 0 to 5 MPag.

From the data presented in **Figure 5**, the relative viscosities are non-Einsteinian and consistent from 4 to 10 °C. As O-GNF loading increases, the ranges of relative viscosity values also increase: 0.7 to 0.8 for 0.1 ppm, 0.8 to 0.9 for 1 ppm, and 0.9 to 1 for 10 ppm. These are mostly lower than the unpressurized range with only air (0.95 to 1). Therefore, the results again suggest that the O-GNFs bring more methane into the system, which would lower the effective viscosity, and that they could add a slight pressure-dependence to the viscosity despite methane's low water solubility. The increase in the relative viscosity ranges observed with loading may simply be due to similar small increases in the internal friction of the nanofluid which are greater than the increase in NE effects. One can also observe a significant rise in relative viscosity at 2 °C for all concentrations. This increase brings the relative viscosity well above 1.00 (Einsteinian regime), which could be connected to temperature-density effects in liquid water. As water cools, its density increases, but at 4 °C, there is an anomalous local maximum, after which density begins to decrease even as cooling continues.[75] Tanaka (2000) proposed that as water cools, there is an ordering transition from density ordering, where the distance between nearest-neighbour molecules is minimized, to bond ordering, where the quality of the hydrogen bonds is maximized for energetic favourability.[75] The latter ordering has bond structures that are similar to those found in ice, which exhibits a lower density compared to liquid water.[76] Additionally, during the transition (at approximately 4 °C), the relative viscosity could increase as there are more ice-like structures present in the water and water containing such structures has previously been measured to have a higher viscosity. However, this increase appears to occur only at lower temperatures. The density of water is mainly due to the strength of its hydrogen bond network, and a 2% decrease in bond strength is sufficient to eliminate the density anomaly (i.e., the curve of density with temperature for water is shifted 4 °C to the left).[77] Viscosity is equally tied to water's hydrogen bond strength: for instance, a 3% increase in hydrogen bond strength can result in a 23% increase in viscosity.[77] As discussed, it is possible that O-GNFs weaken the hydrogen bond network; a bond strength reduction of less than 2% could significantly lower viscosity and shift the density anomaly, and thus the relative viscosity bump, leftward. Therefore, the transition to ice-like structures in water may only occur at lower temperatures in O-GNF nanofluids, due to their ability to disrupt hydrogen bonding, such that a significant increase in relative viscosity occurs at 2 and not 4 °C.

However, it may then be expected that the relative viscosity should increase at 0 °C as the water structure becomes more ice-like, but there is a decrease instead. Note that only the 0.1 and 1 ppm values return to the non-Einsteinian regime and that these values are still greater than what was measured from 4 to 10 °C. As the water transitions further into the bond ordering state, the likelihood of the formation of more ice-like structures increases. However, the fluid is also under shear, which can cause these structures to become distorted and dissociate. Around the 0 to 2 °C range, there may be a discrepancy between the formation of ice-like patches of water and shear stress. While more of these patches could be present at 0 °C, it is possible that the shear rate is so great as to result in more dissociation compared to any dissociation present at 2 °C. Put simply, for the shear rate used in this study, the shear-induced dissociation rate at 0 °C may be larger than that at 2 °C, which would result in a relative viscosity decrease. Moreover, the relative viscosity would still be higher than at 4 °C as some ice-like structures would still be present. However, further investigation would be necessary to confirm these effects as this was not measured as part of this study.

### 3.2.2 Liquid to Solid Phase Transition

Repeating the temperature range, pressure and temperature effects were examined for the three O-GNF concentrations (0.1, 1, and 10 ppm) at pressures from 10 to 30 MPag: these were the hydrate-forming conditions. **Figure 6** shows the 1 ppm O-GNF results for the temporal viscosity evolution of systems that successfully formed hydrates. The same figures for the other concentrations (0.1 and 10 ppm) are found in the Appendix. For all concentrations, the transition from the liquid to solid (hydrate) phase was characterized by an increase in viscosity over time until a maximum was reached where nearly no liquid remained. This maximum was approximately 1200 mPa·s, though it could vary due to the torque limit imposed to protect the measurement device. For instance, many of the 30 MPag runs exhibited lower maxima than their lower-pressure counterparts, as the torque limit was exceeded prior to the subsequent data point measurement. Also, note that the viscosity values could be unstable and vary significantly over short periods. There were three stages to the phase transition. These were the initial growth stage, followed by a hydrate slurry and a final growth stage. In the 10 MPag runs (blue), the presence of these stages is the most apparent as driving forces are low. In the initial growth stage, there was a significant, sudden increase in the measured viscosity. Soon after, the viscosity stabilized as the growth rate became restricted. This behavioural shift marked the beginning of the slurry phase. The slurry consisted of O-GNFs and hydrate clusters suspended in liquid water, so its length was driving force-dependent. The end of the slurry phase was evident when a substantial viscosity rise occurred: this also marked the beginning of the final growth stage, where viscosity increased to a maximum value.[24] The three stages were detected to some extent in all hydrate-forming runs, though some driving forces could be sufficiently large such that the slurry phase was undetectable due to its short length. From **Figure 6**, it is evident that the temporal viscosity behaviour had a correlation to changes in temperature or pressure. Hydrate formation occurred more quickly at higher driving forces (high pressure combined with low temperature), limiting both the slurry phase length and overall time to reach the maximum viscosity. This effect is evident as the horizontal axis limits in the 1 ppm system quadruple with increasing temperature: most runs were complete in about two minutes at 0 °C but required about 12 minutes at 8 °C. Therefore, the shortest phase transitions were observed at 0 °C and 30 MPag, whereas the longest were observed at 8/10 °C with pressures of either 10 or 15 MPag. These same stages and behaviour were also found in the baseline study, so the presence of O-GNFs may only affect the time lengths of the stages. Further discussion is found in the next section.

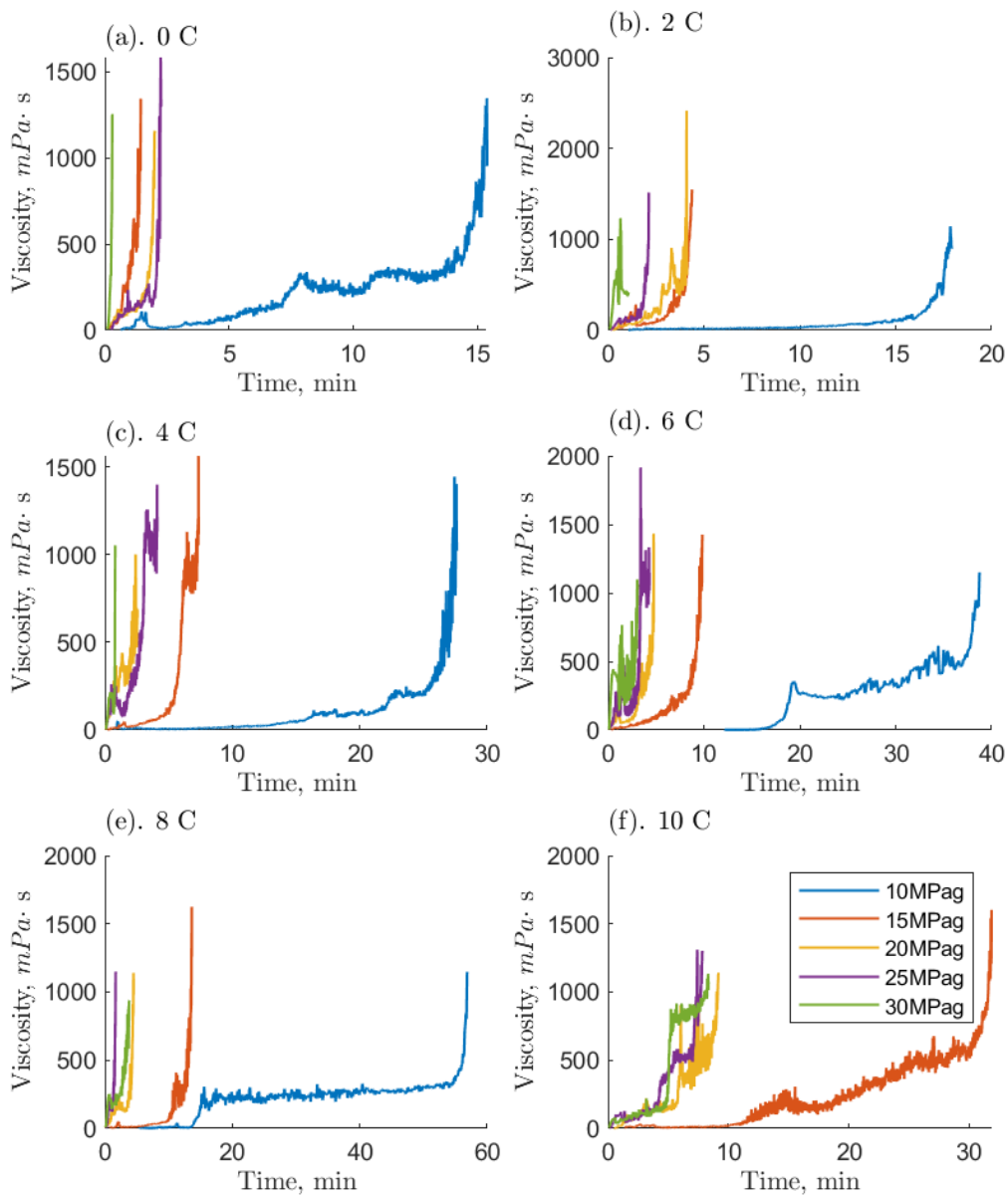

**Figure 6.** Temporal viscosity evolution in the 1 ppm O-GNF-methane-water system under hydrate-forming conditions. Note that time zero is the start of hydrate formation.

As previously mentioned, all systems containing O-GNFs exhibited shorter (1) times to reach the maximum viscosity and (2) slurry phases compared to the baseline. Furthermore, fewer test runs exhibited prominent or prolonged slurry phases in nanoparticle systems. O-GNFs have previously been shown to promote hydrate formation through multiple effects. As described earlier, the shuttle effect would result in greater gas availability to the system, which would increase the mass transfer coefficient.[78] The presence of O-GNFs would also enhance the available gas-liquid interfacial area, though it can be noted that it is unlikely that they act as heterogeneous nucleation sites.[79] They could equally act as microscopic stirrers, creating localized fluid displacements. These mass transfer improvements are considered the most significant for superior nanofluid kinetics, a full order of magnitude more effective than any heat transfer improvements they may provide.[80, 81] Additionally, the O-GNFs can interact with the gas-liquid interface, inducing further turbulence and thinning the diffusion layer for methane. This would also enhance mass diffusivity in the system and increase the hydrate growth rate.[78] It is also possible that the motion of O-GNFs could break up hydrate clusters, which would increase the overall surface area from which hydrates could grow and limit the length of the slurry phase.

Qualitatively, the systems containing 0.1 ppm O-GNF showed the most prolonged slurry phases. The 1 ppm systems have very few runs with a slurry phase and, excluding the lowest pressure runs at each temperature, much shorter slurry phases before the final growth stage was initiated. The 10 ppm systems exhibited more runs with a prominent slurry phase, though less than systems with 0.1 ppm O-GNF. Furthermore, the runs at 0.1 and 10 ppm have a greater separation (i.e., they do not overlap significantly at each temperature condition) than those at 1 ppm. The runs at 1 ppm have significant overlap at the highest pressures and so the highest driving forces, indicating that a thermodynamically limited maximum possible rate was approached in these systems starting at pressures as low as 15 MPag. However, more investigation is required as this may be a characteristic of systems with low initial liquid volumes. Regardless, these results indicate that the greatest increases in growth rate occurred at the 1 ppm concentration. Previous studies have also found that loadings of 1 ppm O-GNF showed the greatest methane hydrate growth rates, followed by 10 and 0.1 ppm systems.[21] As the concentration increased from 0.1 to 1 ppm, the surface and mixing effects were likely increased, improving conditions for growth. However, it is possible that when the concentration was further increased, the average distance between nanoparticle collisions, the mean free path, was reduced. This could impede the free motion of the O-GNFs and counteract improvements in mass diffusivity.[81] It could also lower the effective gas-liquid interfacial area; while the nanoparticles likely do not agglomerate, the number of times at which they momentarily come into contact before repulsion increases with concentration, limiting the amount of O-GNF surface area participating in system effects.[22] These could account for the apparent reduction in growth rate at 10 ppm, which will be discussed quantitatively in the next section.

One notable hydrate-forming run was the 1 ppm 8 °C/10 MPag condition (4.1 MPa driving force), which did not form hydrates in the required 90-minute period for the baseline or in the 0.1 nor 10 ppm systems. It is the lowest driving force condition at which methane hydrates have formed in a high-pressure rheometer system to date. This may further indicate that the system is most enhanced when loaded with 1 ppm O-GNF. However, this also means that nine lower driving force conditions (from 0.2 to 2.7 MPa, still relatively high) did not form hydrates at any concentration in the allotted time though they have resulted in the successful, rapid formation in many other systems previously.[21, 23] High-pressure rheological instrumentation, which is required to measure the viscosity of hydrate-forming systems, often comes with certain limitations. As mentioned previously, the rheometer induces a high-shear environment that could cause mechanical dissociation of gas hydrate nuclei during the nucleation stage of formation before they reach a critical radius. The lack of directionality in the vorticity of the shear flow combined with extensional effects that are ± 45 ° to the flow direction could separate neighboring groups of nuclei and reduce the effective driving force. Furthermore, the smooth stainless-steel surfaces of the well and measurement system, as well as the low number of impurities in the RO water, may provide few sites for hydrate nucleation. Moreover, the system could impose diffusion limitations on hydrate formation due to the small sample volume and double annulus measurement geometry. Therefore, there would be a low gas-liquid surface area for mass diffusion, and the dissolution of gas into the liquid phase would be limited.[24] However, O-GNFs have previously been measured to enhance the rates of methane gas dissolution in water by about 45%[22]. Therefore, this may not be a significant factor considering only a single condition at a single concentration was measured to be affected. Lastly, hydrate formation is exothermic, and sufficient heat evolution can result in the system becoming self-limiting, notably creating less favorable nucleation conditions.[4] While O-GNFs may improve heat dissipation in the system, they also promote hydrate growth, so the rate of heat generation is also increased: the temperature in O-GNF systems compared to the baseline may not have changed significantly. Therefore, the addition of O-GNFs largely did not overcome the limitations inherent to the measuring system, and the conditions at which hydrates formed (within the 90-minute timeframe) did not change significantly from the baseline.

These limitations are likely strong influences on the slurry phase, which, as discussed, was shorter in O-GNF systems. Viscosity could drop over time at intermediate conditions as higher temperatures resulted in some hydrate dissociation. However, this occurred less frequently in systems with O-GNFs and the times for the viscosity to recover were also reduced. This suggests that O-GNFs do not greatly reduce the limitations of the system prior to hydrate formation but have some effect during growth. As hydrates formed and more methane was used, the enhanced amount of methane and improved mixing brought about by the O-GNFs may have reduced limitations to mass transfer, so the system maintained faster growth rates and higher viscosities.

### 3.2.3 Methane Hydrate Growth Kinetics for Applications

Hydrate technologies ideally do not reach near-solidification viscosities, so the times required to reach 200 ($T_{200}$) and 500 ($T_{500}$) mPa·s are presented in **Figure 7** for the 1 ppm system. The figures for the other O-GNF concentrations (0.1 and 10 ppm) are found in the Appendix. Higher viscosity values (than those selected) are often accompanied by significant pumping requirements; technological applications of hydrates may limit the amount by which viscosity increases. The time to reach maximum viscosity, around 1200 mPa·s, is less relevant to process design and the additional time for the system to reach viscosities greater than 500 mPa·s was negligible. The runs with the lowest driving forces have the longest times, and those with the highest driving forces are significantly faster. However, as the driving force increases, $T_{200}$ and $T_{500}$ value decreases are not linear. Instead, the greatest decrease occurs between 10 and 15 MPag, where the times are cut approximately in half, compared to those at 20 or 25 MPag, which exhibited little change. This behavior is similar to the baseline and was present at all concentrations.[24] However, for the $T_{200}$ and the $T_{500}$ (in particular), the curves did not flatten as much or at as low of pressures in the baseline as they did in this study (see **Figure 7**b). This indicates that the growth rate enhancement provided via the presence of O-GNFs results in very similar times to a desired viscosity between different high driving force pressures. It may also indicate that the system's limits are being approached. This is ideal for technologies employing these nanoparticles as it suggests that the time to achieve a specific viscosity value can be reached with less severe conditions. For instance, the $T_{500}$ in the 1 ppm system at 8 °C/30 MPag was 3.15 minutes, and at 6 °C/20 MPag was 3.27 minutes (compared to 3.98 and 6.57 minutes at these respective conditions in the baseline). These are relatively similar times to the same viscosity (i.e., for just a 2 °C lowering of the temperature, one could get the same result at a pressure that is 10 MPa lower). In other words, even though the driving force was lowered by 8.8 MPa between these systems, the times to 500 mPa·s are similar. This could positively affect critical design considerations for hydrate technologies, such as cost and operation.

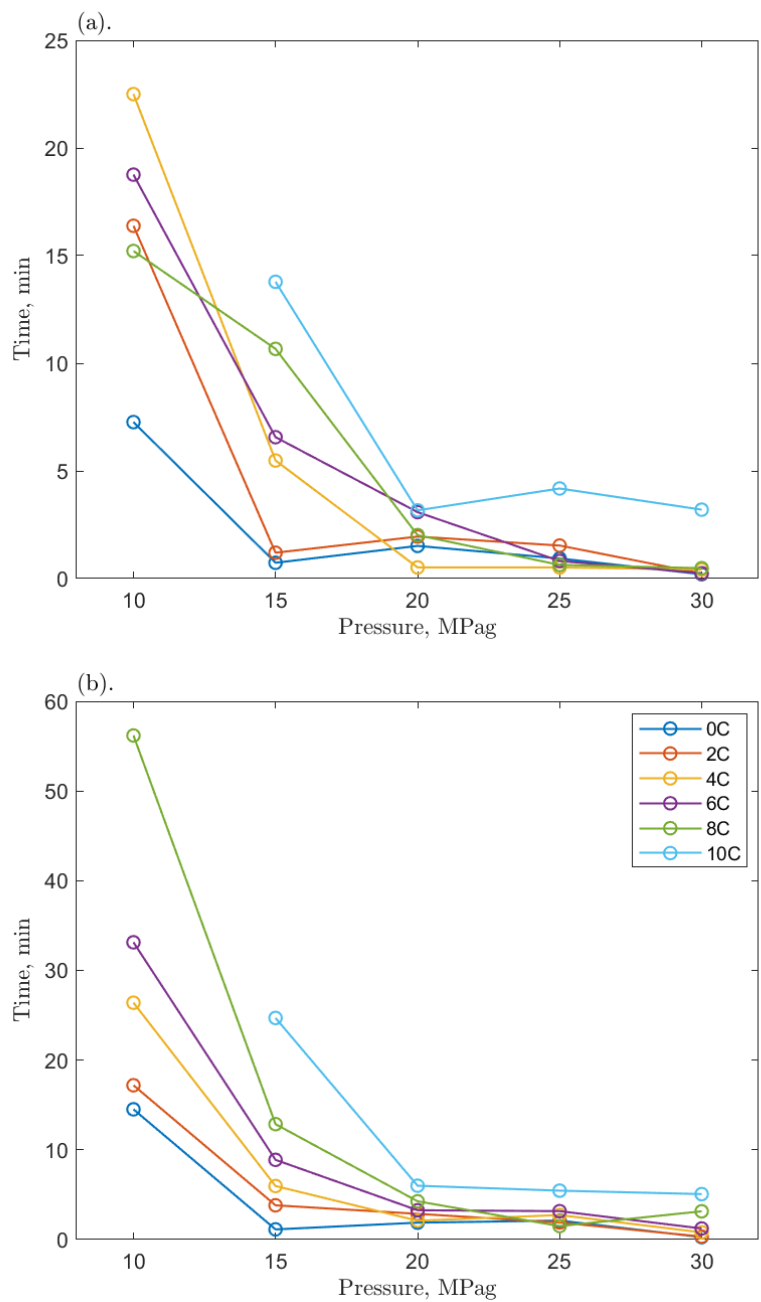

**Figure 7.** Isotherms that illustrate the time for the 1 ppm O-GNF-methane-water system to reach (a) 200 mPa·s and (b) 500 mPa·s from the beginning of hydrate formation.

The $T_{200}$ and $T_{500}$ values for the O-GNF systems were reduced, indicating a faster growth rate compared to the baseline. Across the temperature, pressure, and concentration ranges, no distinct trend was observed regarding changes from the baseline. However, there were clear effects when average values were taken across all conditions at each concentration. These averages consist of 28 temperature/pressure conditions, though their significance is likely low because they are so generalized. It is possible that this study's short time frames, with some runs being complete in a matter of seconds and others in several minutes, and the stochasticity of nucleation, caused this lack of trends. The $T_{200}$ times were still reduced by 25.78, 49.75, and 10.53 % on average for the 0.1, 1, and 10 ppm systems, respectively. The $T_{500}$ times were reduced by 11.36, 31.93, and 6.02 % on average for the 0.1, 1, and 10 ppm systems, respectively. The 1 ppm system had the most significant reduction in both cases and was thus the fastest system, followed by 0.1 and 10 ppm. This may contradict the results shown in **Figure 6** and the accompanying figures in the Appendix, which showed that the 10 ppm system was faster than the 0.1 ppm system. This is due to the over-generalization of the average values. While the 10 ppm system is generally faster than the 0.1 ppm system under most conditions, at lower driving forces, it had more values closer to the baseline than the 0.1 ppm system, which brought down the overall average. Moreover, the 10 ppm systems tended to reach the maximum viscosity faster than the 0.1 ppm system, but could be slower to 200 or 500 mPa·s. This could show that enhancement (time reduction) in the 10 ppm system was more beneficial later in hydrate formation but would also be less beneficial than the 0.1 ppm system for reaching viscosity values relevant to technological applications. It would also explain why the 0.1 ppm system has higher enhancement values on average.

The enhancement values are greater for the 200 mPa·s times than those for 500 mPa·s regardless of concentration. This is likely because as the viscosity of the system increases, the mass transfer becomes more limited, even in enhanced systems. Thus, O-GNFs have higher effectiveness in systems with less impendence to mass transfer (i.e., those below 200 mPa·s) because mass transfer improvements are the primary mechanism by which they promote hydrate growth. Previous studies have shown that hydrate growth rates in a system with only a 1.5 MPa driving force are enhanced by 128.20, 287.99, and 246.95 % at 0.1, 1, and 10 ppm O-GNF, respectively, which are significantly higher values than those measured in this study.[21] However, that study used the first 15 minutes of hydrate growth at a lower driving force, which avoided significant changes in viscosity. Additionally, the geometry was a more open, cylindrical geometry with little shear and not a high-shear annulus, allowing for better mixing and less dissociation of hydrate nuclei. Lastly, the enhancement values here have variable timeframes as viscosity could decrease and lengthen the time to the desired values. Therefore, similar trends could not be expected between these studies, and the level of hydrate kinetic enhancement is significantly reduced.

## 4. CONCLUSIONS

O-GNF nanofluid viscosity was examined for 0.1 to 10 ppm concentrations with pressures of 0 to 30 MPag and temperatures of 0 to 10 °C. This was the first time plasma-functionalized graphene's viscosity had been measured in liquid, high-pressure, or hydrate-forming systems. The concentration range explored in this study was also novel for viscosity measurements. The addition of O-GNFs to the system did not affect how water viscosity depends on temperature. However, instead of the effective viscosity increasing, which is expected behaviour, the viscosity was instead reduced by the presence of the nanoparticles: termed for the first time here as non-Einsteinian (NE) viscosity. This may have been because the O-GNFs were well-dispersed and had high specific surface areas. A solvation layer could form at the hydrophobic portion of the O-GNF surface, where slip effects were also present, reducing the local hydrogen bond strength. In addition, local density fluctuations were enhanced at this surface. Considering free volume theory, there would be an increase in larger empty sites and weaker intermolecular interactions (i.e., there would be less resistance to water diffusing to those sites, and the effective viscosity would be reduced). Internal friction, which would raise viscosity, may be overcome by the accumulation of these surface effects. This is even though the concentrations were ultra-low, as the surface area remained high due to the large O-GNF specific surface area. However, the non-Einsteinian hypothesis was not tested further in this study, so computational modelling work is suggested as future work.

O-GNFs did not affect the pressure-dependence of viscosity in water, except at 10 ppm, where the shuttle effect may have increased the presence of hydrophobic methane bubbles enough to have small reductions in viscosity. When pressurized, the system retained its non-Einsteinian relative viscosity behavior, except at 2 °C temperatures. This could be because the maximum density of water was moved to a temperature colder than 4 °C as the hydrogen bonding network was weaker. The effects of the shear environment on hydrate nuclei formation and deformation, and the presence of ice-like structures, may also have played a role in the anomalous behavior. The phase transition of liquid water to solid hydrate was divided into (1) initial growth or increase in viscosity, (2) an O-GNF-hydrate cluster slurry, and (3) a final, rapid viscosity increase to a maximum value. Compared to the baseline, the times to that maximum in O-GNF systems were much shorter. Moreover, due to mass transfer enhancement, the times in which the system exhibited slurry phase behavior were reduced. The 1 ppm O-GNF system was the fastest, followed by the 10 and 0.1 ppm systems. The 10 ppm system may not have been the fastest due to possible limitations of the mean free path at that concentration. In comparison with the baseline study, the addition of O-GNFs, with only a single exception, did not overcome hydrate formation limitations inherent to the measurement system. The times to reach application-relevant viscosity values were maximally 49.75 % (for 200 mPa·s) and 31.93 % (for 500 mPa·s) faster than the pure water/methane baseline. Under this measurement, the 0.1 ppm system was faster than the 10 ppm system, even though the latter achieved the maximum viscosity sooner. This was because the enhancement values were generalized over the entire range of conditions, and the 10 ppm system had more $T_{200}$ and $T_{500}$ values closer to the baseline. Furthermore, the $T_{200}$ enhancement was greater than that for the $T_{500}$ as there were fewer mass transfer limitations at lower viscosities. Therefore, the presence of O-GNFs enhanced hydrate formation, allowing for shorter times to desired viscosities and at lower driving forces than the baseline. NE effects may also make them ideal for heat transfer fluids, and future work is recommended for such applications at the concentrations examined in this study.

AUTHOR INFORMATION

**Corresponding Author**

*phillip.servio@mcgill.ca

**Author Contributions**

The manuscript was written through the contributions of all authors. All authors have given approval to the final version of the manuscript.

ACKNOWLEDGEMENTS

The authors would like to acknowledge the financial support from the Natural Sciences and Engineering Research Council of Canada (NSERC) and the Faculty of Engineering of McGill University (MEDA, Vadasz Scholars Program).

APPENDIX

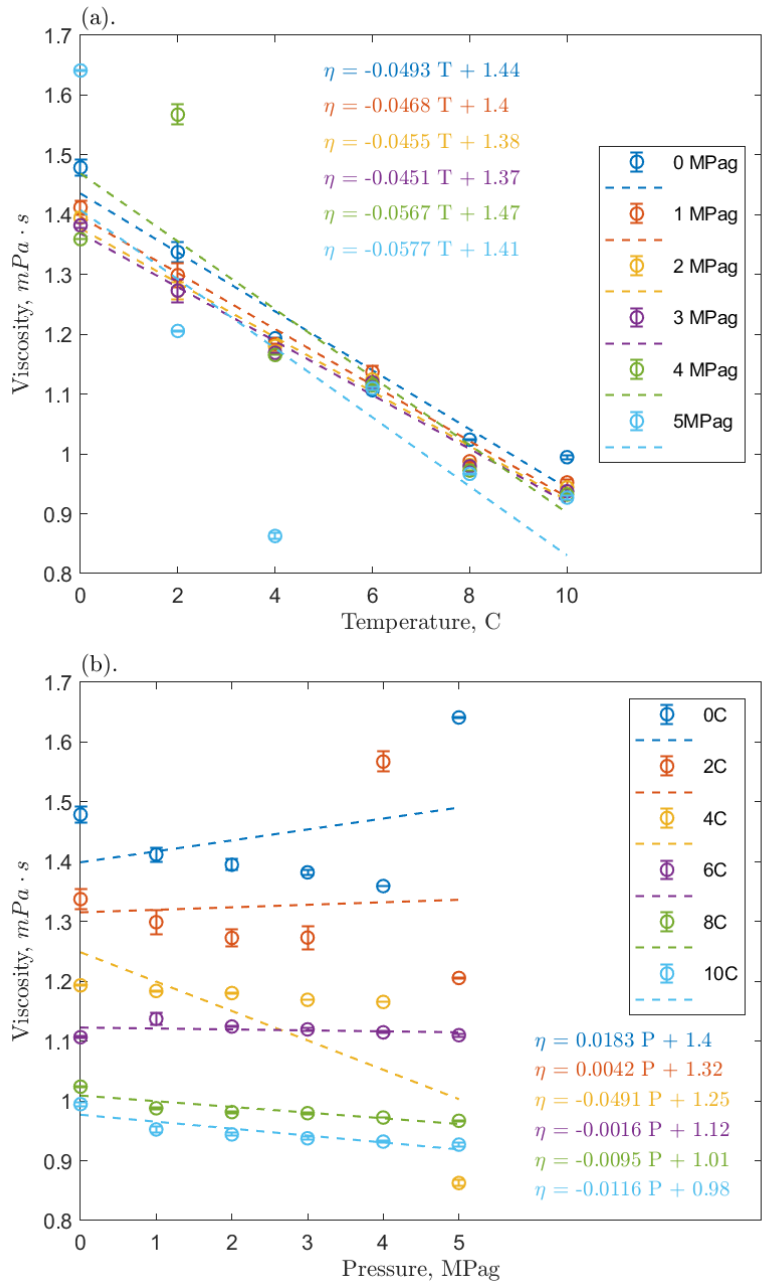

**Figure A1.** Viscosity effects in the 0.1 ppm O-GNF-methane-water system of (a) temperature and (b) pressure. The error bars are 95% confidence intervals and linear regressions are provided for each condition.

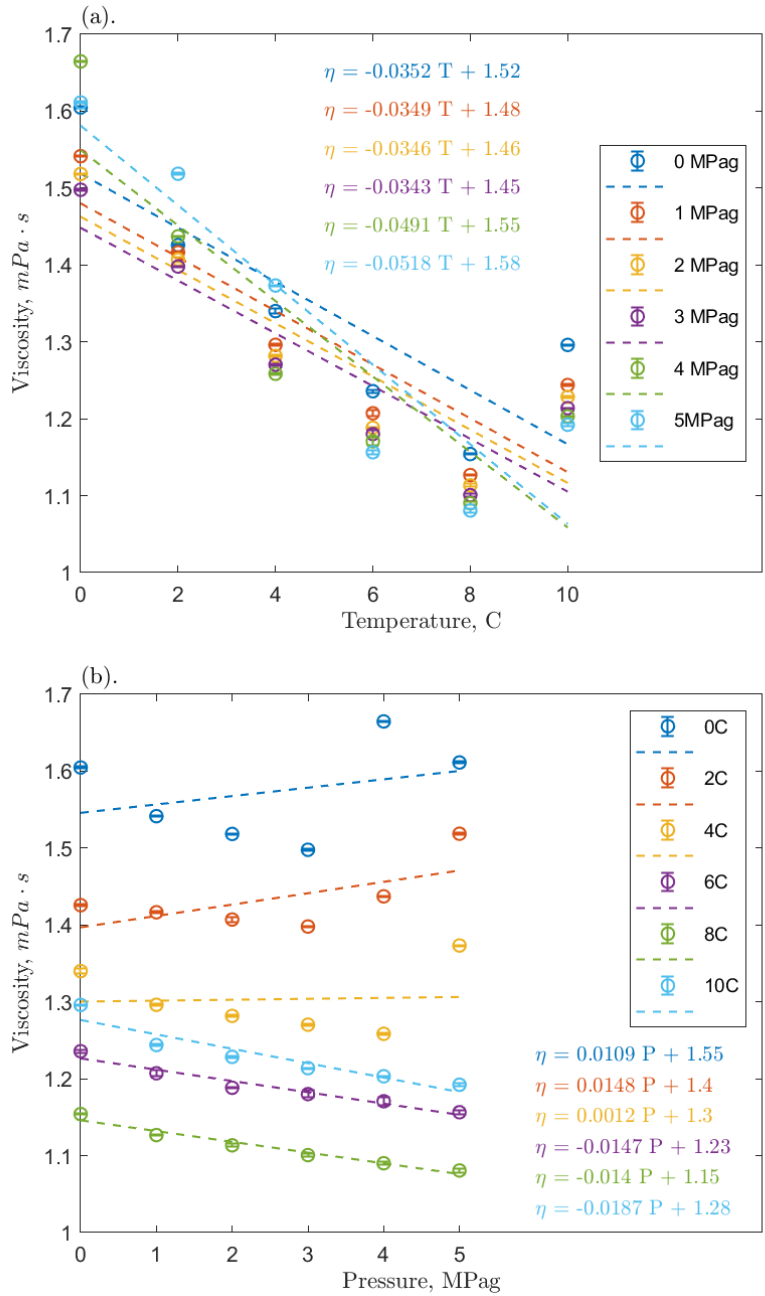

**Figure A2.** Viscosity effects in the 1 ppm O-GNF-methane-water system of (a) temperature and (b) pressure. The error bars are 95% confidence intervals and linear regressions are provided for each condition.

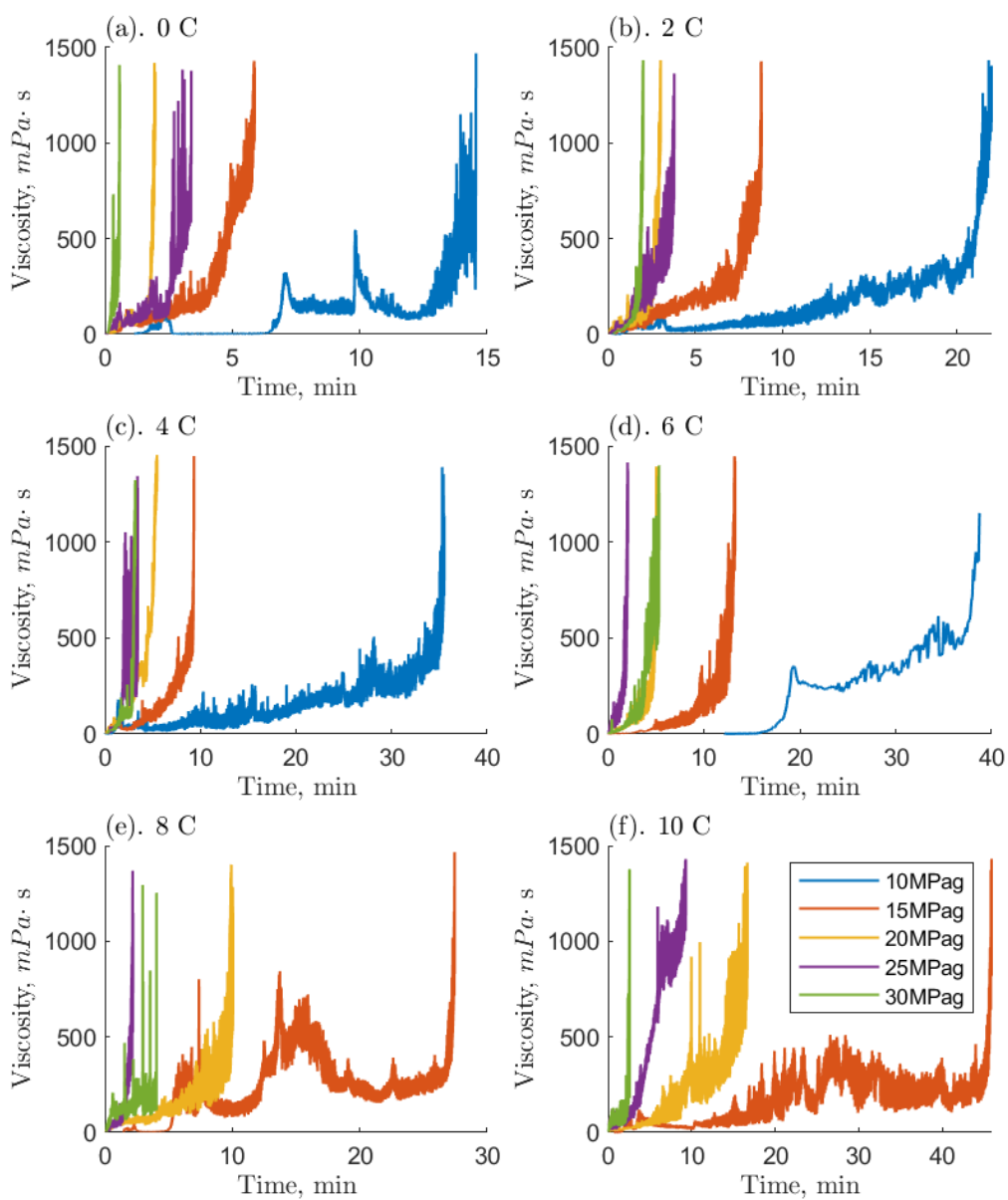

**Figure A3.** Temporal viscosity evolution in the 0.1 ppm O-GNF-methane-water system under hydrate-forming conditions. Note that time zero is the start of hydrate formation.

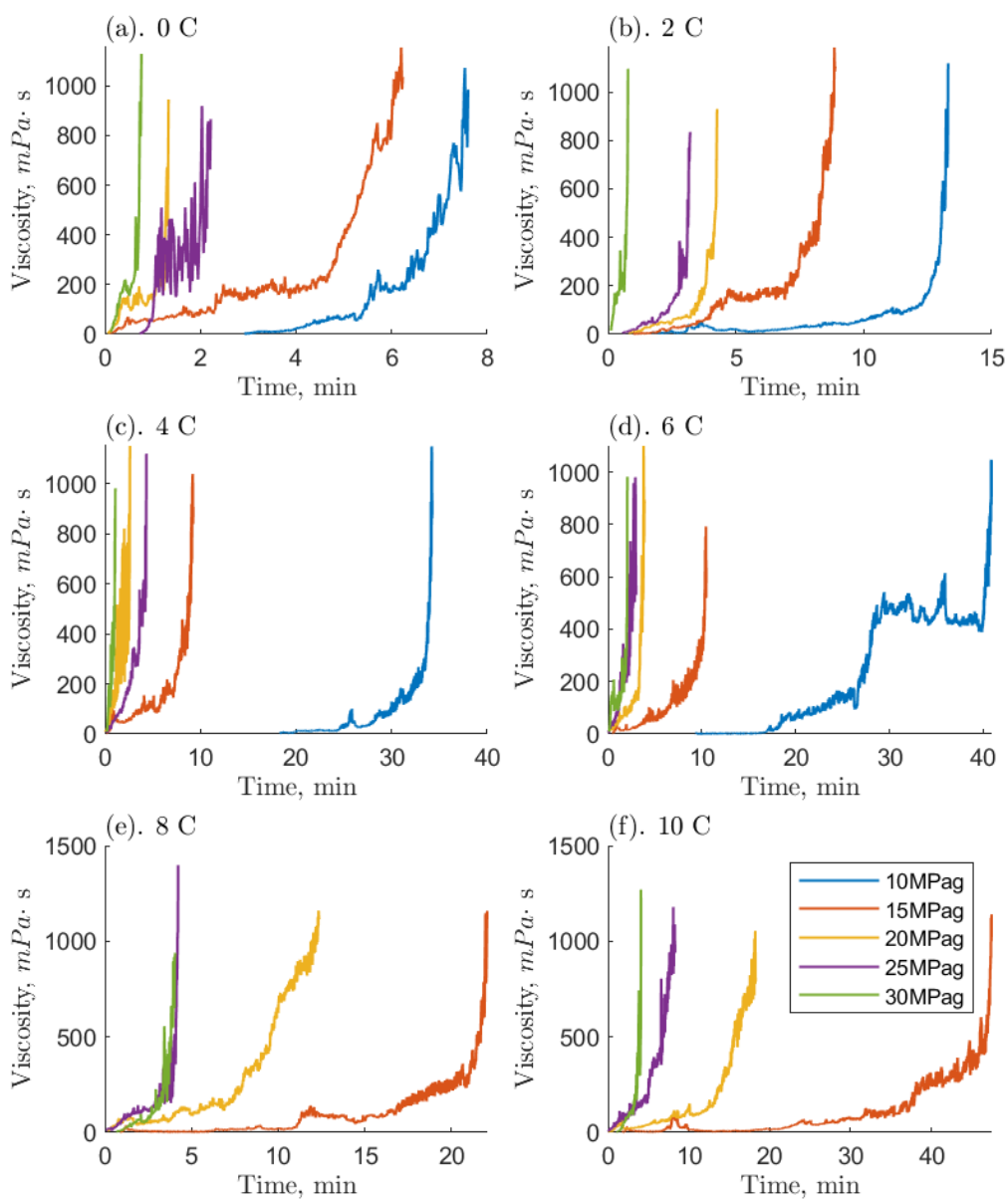

**Figure A4.** Temporal viscosity evolution in the 10 ppm O-GNF-methane-water system under hydrate-forming conditions. Note that time zero is the start of hydrate formation.

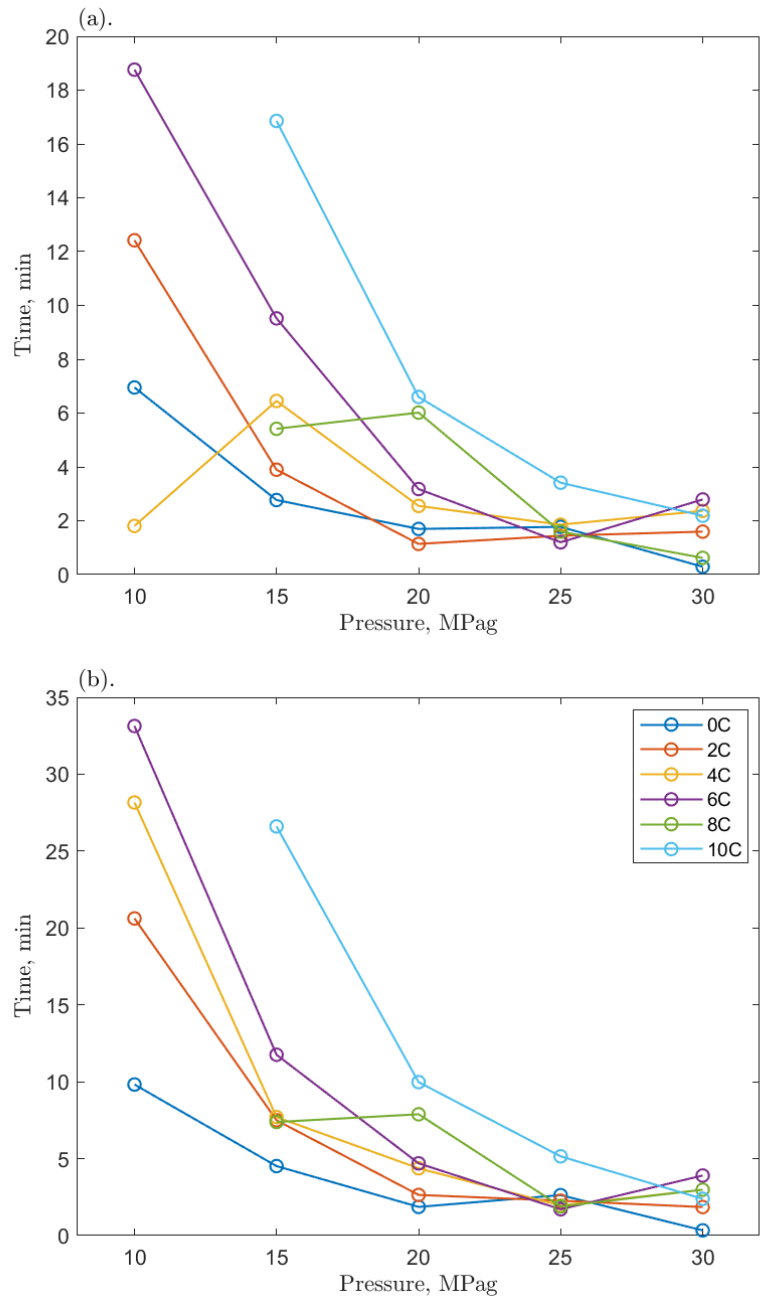

**Figure A5.** Isotherms that illustrate the time for the 0.1 ppm O-GNF-methane-water system to reach (a) 200 mPa·s and (b) 500 mPa·s from the beginning of hydrate formation.

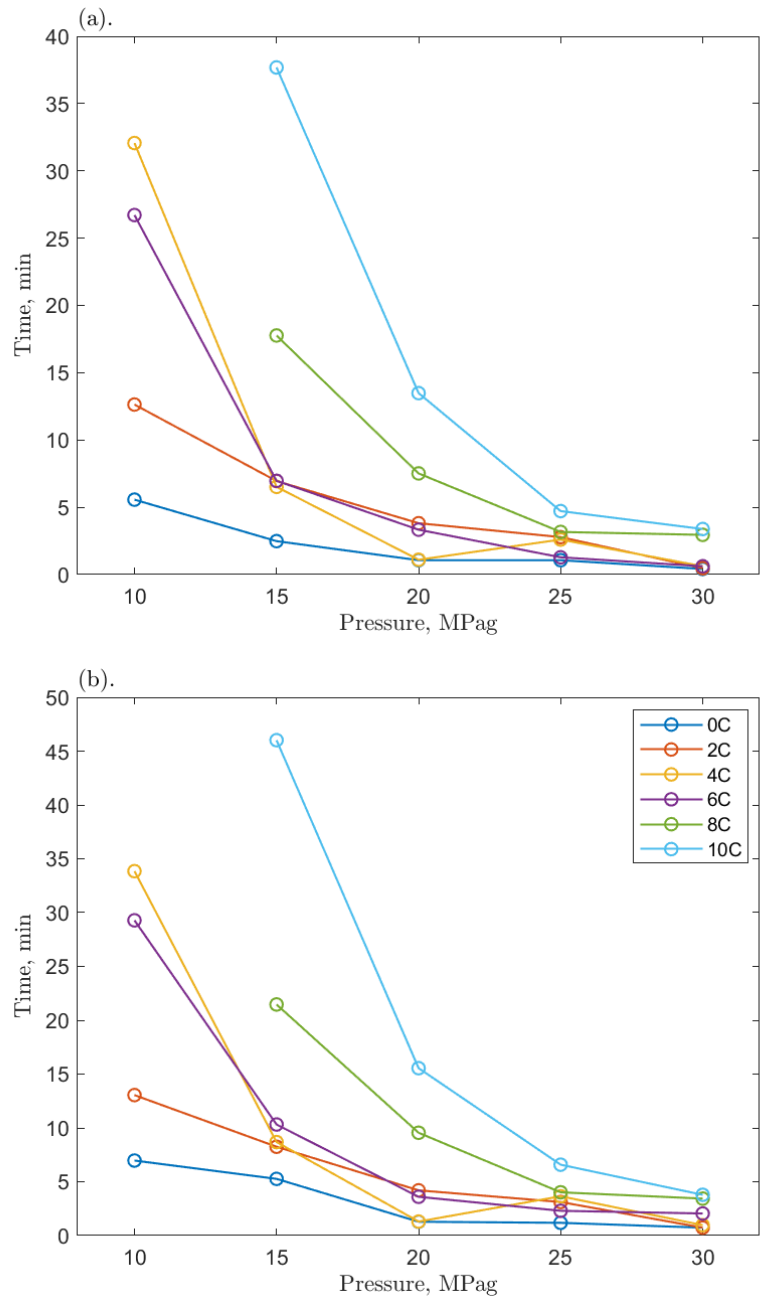

**Figure A6.** Isotherms that illustrate the time for the 10 ppm O-GNF-methane-water system to reach (a) 200 mPa·s and (b) 500 mPa·s from the beginning of hydrate formation.